\documentclass[twocolumn,showpacs,aps,prl,superscriptaddress,USletter]{revtex4}
\usepackage{graphicx}
\usepackage{dcolumn}
\usepackage{amsmath}
\usepackage{epsfig}

\input pubboard/babarsym.tex

\newcommand{\bflav}{\ensuremath{\B_{{\rm flav}}}}

\newcommand{\BABARPubYear}    {05}
\newcommand{\BABARPubNumber}  {017}

\newcommand{\SLACPubNumber} {11247}

\def\DeltaEStd  {\ensuremath{\Delta E} \xspace}

\newcommand{\BZ}{\mbox{${B}^0$}}
\newcommand{\BZBAR}{\mbox{$\overline{B}^0$}}

\def\DpDm{\ensuremath{D^{+} D^{-} }}

\def\DstpDm{\ensuremath{D^{*+}D^{-}}}
\def\DstmDp{\ensuremath{D^{*-}D^{+}}}
\def\Dstpm{\ensuremath{D^{*\pm}}}
\def\Dpm{\ensuremath{D^{\pm}}}
\def\DbothstpmDmp{\ensuremath{D^{(*)\pm}D^{\mp}}}

\def\BDpDm{\ensuremath{\BZ \rightarrow D^{+}D^{-}}}
\def\BDstpDm{\ensuremath{\BZ \rightarrow D^{*+}D^{-}}}
\def\BDstmDp{\ensuremath{\BZ \rightarrow D^{*-}D^{+}}}

\def\BDstpmDmp{\ensuremath{\BZ \rightarrow D^{*\pm}D^{\mp}}}
\def\BDbothstpmDmp{\ensuremath{\BZ \rightarrow D^{(*)\pm}D^{\mp}}}

\def\sss{\scriptscriptstyle}
\def\barpd{{\raise.35ex\hbox{${\sss (}$}}--{\raise.35ex\hbox{${\sss )}$}}}
\def\BorBbar{\hbox{$B$\kern-0.85em\raise1.5ex\hbox{\barpd}\hspace{-0.4mm}$^0$}}

\def\BorBbarDpDm{\ensuremath{\BorBbar \rightarrow D^{+}D^{-}}}
\def\BorBbarDstpDm{\ensuremath{\BorBbar \rightarrow D^{*+}D^{-}}}
\def\BorBbarDstmDp{\ensuremath{\BorBbar \rightarrow D^{*-}D^{+}}}

\def\masslik{\ensuremath{\cal{L}_{\rm mass}}}

\def\figurebox#1#2#3{%
    \def\arg{#3}%
    \ifx\arg\empty
    {\hfill\vbox{\hsize#2\hrule\hbox to #2{\vrule\hfill\vbox to #1{\hsize#2\vfill}\vrule}\hrule}\hfill}%
    \else
    {\hfill\epsfbox{#3}\hfill}%
    \fi}

\begin{document}

\preprint{\babar-PUB-\BABARPubYear/\BABARPubNumber}
\preprint{SLAC-PUB-\SLACPubNumber}

\begin{flushleft}
\babar-PUB-\BABARPubYear/\BABARPubNumber\\
SLAC-PUB-\SLACPubNumber\\
\end{flushleft}

\title{
{\large \boldmath \bf  Measurement of Time-Dependent \CP Asymmetries in \BDbothstpmDmp\ Decays}
}

%
\author{B.~Aubert}
\author{R.~Barate}
\author{D.~Boutigny}
\author{F.~Couderc}
\author{Y.~Karyotakis}
\author{J.~P.~Lees}
\author{V.~Poireau}
\author{V.~Tisserand}
\author{A.~Zghiche}
\affiliation{Laboratoire de Physique des Particules, F-74941 Annecy-le-Vieux, France }
\author{E.~Grauges}
\affiliation{IFAE, Universitat Autonoma de Barcelona, E-08193 Bellaterra, Barcelona, Spain }
% ...
\collaboration{The \babar\ Collaboration}
\noaffiliation

\date{\today}

\begin{abstract}
We present a first measurement of \CP asymmetries in neutral $B$ decays to \DpDm, and updated
\CP asymmetry measurements in decays to \DstpDm\ and \DstmDp.
We use fully-reconstructed decays collected in a data sample 
of $(232 \pm 3) \times 10^6$ $\FourS\to B\Bbar$
events in the \babar\ detector at the
\pep2\ asymmetric-energy \BF\ at SLAC.
We determine the time-dependent asymmetry parameters to be
$S_{\DstpDm} = -0.54 \pm 0.35 \pm 0.07 $,
$C_{\DstpDm} = 0.09 \pm 0.25 \pm 0.06$,
$S_{\DstmDp} = -0.29 \pm 0.33 \pm 0.07$,
$C_{\DstmDp} = 0.17 \pm 0.24 \pm 0.04$,
$S_{\DpDm} = -0.29 \pm 0.63 \pm 0.06$, and
$C_{\DpDm} = 0.11 \pm 0.35 \pm 0.06$, where
in each case the first error is statistical and the second error is systematic.
\end{abstract}


\maketitle
Charge-parity (\CP) violation is described in the Standard Model (SM) by a single irreducible complex phase in the
Cabibbo-Kobayashi-Maskawa (CKM) quark mixing matrix $V$~\cite{CKM}. The $B$-meson system provides an excellent probe
for testing the completeness of the CKM mechanism in a variety of \CP asymmetries~\cite{BCP}. Measurements of \CP violation
in $\Bz \to (\ccbar) K^{0(*)}$ decays~\cite{chargeconj} by the \babar~\cite{babar-stwob-prl} and Belle~\cite{belle-stwob-prl} collaborations 
have precisely determined the parameter \stwob, where $\beta$ is $\arg \left[\, -V_{\rm cd}^{}V_{\rm cb}^* / V_{\rm td}^{}V_{\rm tb}^*\, \right]$. 
The current world average of $\stwob = 0.726\pm0.037$ is in good agreement with the range implied by 
other measurements in the context of the SM~\cite{CKMfitters}, providing evidence that the CKM mechanism is the main source of \CP violation in the quark sector.

Decays of \BZ\ mesons to pairs of charged $D^{(*)}$ mesons can also be used to determine \stwob.
These decays proceed to leading order via a tree-level color-allowed $b \to c\bar{c}d$ transition.
The presence of a gluonic penguin contribution with a different weak phase is expected to change the
magnitude of the \CP asymmetry by not more than a few percent~\cite{dstdexpect}. However, additional 
contributions from non-SM
processes may lead to shifts as large as  $\Delta \beta \approx 0.6$ in some models~\cite{GrossWorah}.
Interference between SM penguin and tree amplitudes can additionally provide some sensitivity to the angle
$\gamma = \arg \left[\, -V_{\rm ud}^{}V_{\rm ub}^* / V_{\rm cd}^{}V_{\rm cb}^*\, \right]$~\cite{DDgam}.  

In this Letter we present a first measurement of \CP asymmetries in the decay \BDpDm\ and 
improved measurements of \CP asymmetries 
in \BDstpDm\ and \BDstmDp\ decays~\cite{oldbabarDstD,belleDstD}. 
The results are based on an analysis of $(232 \pm 3) \times 10^6$ $\FourS\to B\Bbar$ decays recorded by
the \babar\ detector~\cite{babar-detector-nim} at the PEP-II  $e^+e^-$ collider.

The selection of \BDstpDm\ candidates is similar to that of our previous analysis~\cite{oldbabarDstD}.
We reconstruct \Dstarp\ in its decay to $\Dz\pip$, 
where the \Dz is reconstructed in one of four final states: $K^{-}\pi^{+}$, $K^{-}\pi^{+}\piz$, $K^{-}\pi^{+}\pi^{-}\pi^{+}$, 
or $\KS\pi^{+}\pi^{-}$. The \Dm\ is reconstructed in the final states $K^{+}\pi^{-}\pi^{-}$ or $\KS\pi^{-}$.
The \KS candidates are reconstructed from $\pi^{+}\pi^{-}$ pairs within 15~\mevcc of the nominal \KS\ mass~\cite{PDG2004}. 
The \piz candidates are reconstructed as photon pairs with an invariant mass between 115 and 150~\mevcc;
each photon must have energy above 30~\mev in the laboratory frame and the sum of the photon energies must exceed 200~\mev.
We require the \Dz\ and \Dpm\ candidates to have reconstructed invariant masses within
20~\mevcc of their respective nominal masses, except for \Dz decays with a \piz daughter,
which must be within 35~\mevcc of the nominal \Dz mass.
The \BDpDm\ candidates are reconstructed solely through the decay of $D^{\mp} \to K^{\pm}\pi^{\mp}\pi^{\mp}$. 
Charged kaons are required to be incompatible with a pion hypothesis on the basis of measurements of particle identification quantities.

To reduce background from continuum events ($\epem\ra q\qbar,\q = u,d,s,c$) is reduced, we exploit the contrast 
between the spherical topology of \BB\ events and the more jet-like nature of continuum events. 
We require the ratio of the second-to-zeroth order Fox-Wolfram 
moments~\cite{FoxWolfram} to be less than 0.6. We also use a Fisher discriminant, 
constructed as an optimized linear combination of 11 event shape variables~\cite{Fisher}: the momentum flow in nine concentric cones around the 
thrust axis of 
the reconstructed $\BZ$ candidate, the angle between that thrust axis and the beam axis, and the angle between the line-of-flight 
of the $\BZ$ candidate and the beam axis. 
The Fisher discriminant selection requirement increases the signal significance
by 2\% in the case of \BDstpmDmp\ and 9\% in the case of \BDpDm.

\begin{figure*}[!tb]
\begin{center}
\includegraphics[width=0.32\linewidth]{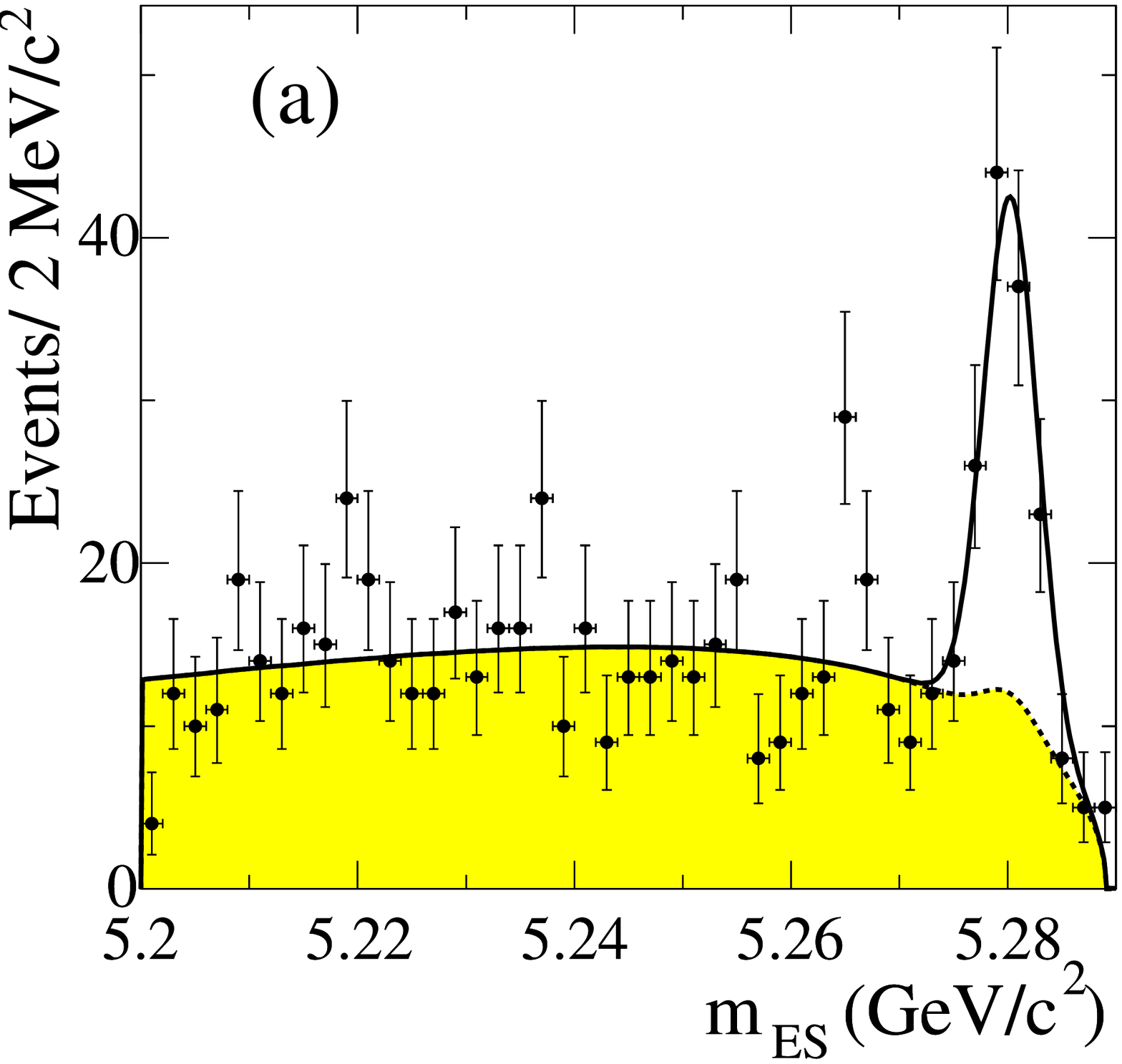}
\includegraphics[width=0.32\linewidth]{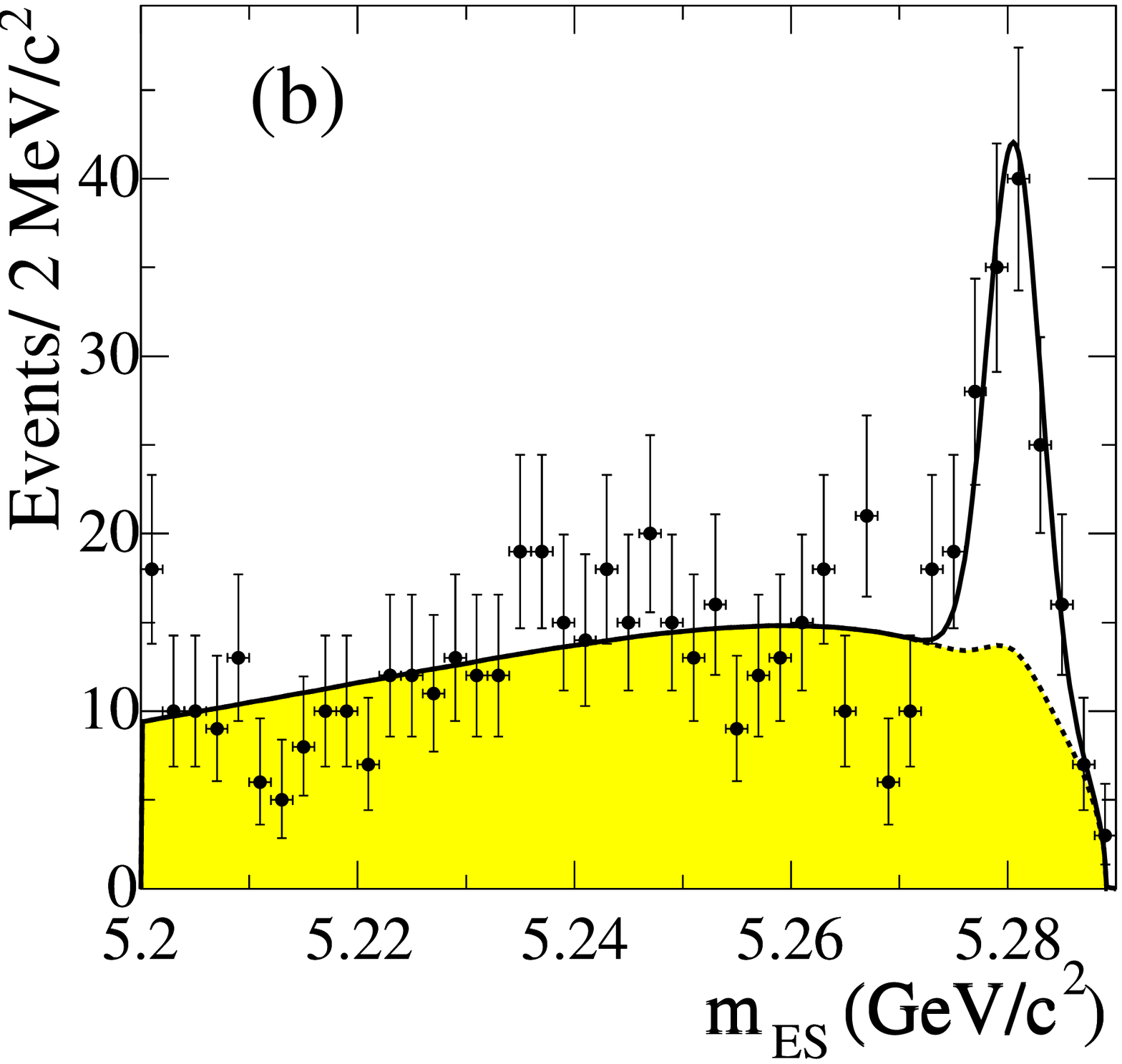}
\includegraphics[width=0.32\linewidth]{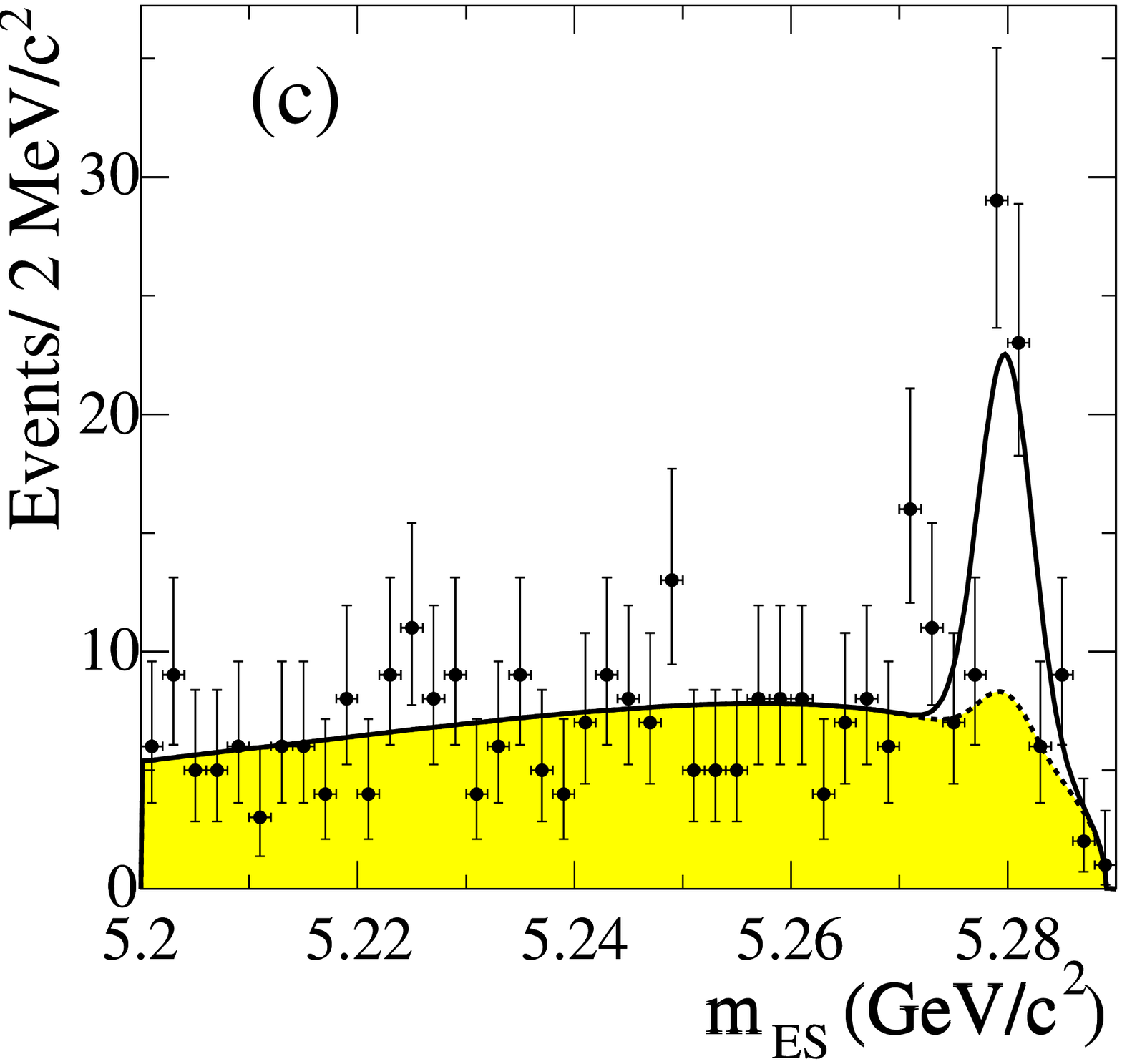}
\end{center}
\caption{Distribution of $\mes$ for (a) \BorBbarDstpDm,  (b) \BorBbarDstmDp\ and (c) \BorBbarDpDm\ candidates. 
The shaded areas represent the contributions from background events.
The dashed and solid curves describing the background and signal plus background distributions 
respectively are explained in the text.
}
\label{fig:mesplots}
\end{figure*}

For each candidate, we construct a likelihood variable \masslik\ from the differences between
the reconstructed masses and the nominal masses of the \Dstarp, \Dp, and \Dz\ candidates~\cite{oldbabarDstD}. 
The \masslik\ variable is the product of the likelihood functions for the three
candidate types. The likelihood for \Dp\ and \Dz\ is parametrized with a single Gaussian function,
while the mass difference  $m_{\Dstarp} - m_{\Dz}$ is parameterized as the sum of two Gaussian functions.
The computed value of \masslik\ and the difference \DeltaEStd\ 
between measured energy of the \BZ\ candidate in the center-of-mass frame and half the center-of-mass energy, 
$\DeltaEStd \equiv E_{B}^* - (\sqrt{s}/2)$, are used to reduce the combinatoric background. Maximum allowed values
for both $-\ln\masslik$ and $|\DeltaEStd|$ are set for each individual final state separately, 
optimized using a Monte Carlo simulation~\cite{Geant4} to obtain the highest expected signal significance.

The technique for measuring the \CP\ asymmetries is analogous to previous \babar\ measurements
described in detail elsewhere~\cite{babar-stwob-prd}. After the reconstruction of a
\BDbothstpmDmp\ candidate $B_{CP}$, we assign the remaining tracks in the event to the other $B$ 
meson $B_{\rm tag}$. We compute a proper time difference $\deltat$ and its estimated uncertainty $\sigma_{\deltat}$ from 
the reconstructed decay vertices of $B_{CP}$ and $B_{\rm tag}$. The tracks assigned to $B_{\rm tag}$ are used 
to determine the $B_{\rm tag}$ flavor and thus the flavor of the $B_{CP}$ meson at $\deltat=0$~\cite{babar-stwob-newprl}. 
Events are classified in one of six tag categories and must have an estimated probability \mistag of assigning the wrong 
flavor to $B_{\rm tag}$ less than 45\%.

Taking into account the uncertainty in the vertex position and tag flavor, the 
observed $\deltat$ distribution for \BDbothstpmDmp\  signal events ${\rm F}_\pm^{CP}(\deltat)$ is described by:
\begin{eqnarray}
{\rm F}_\pm^{CP}(\, \deltat) = {\frac{e^{{- \left| \deltat' \right|}/\tau_{\Bz} }}{4\tau_{\Bz} }} 
 \Bigg\{ 1 \pm  (1-2\mistag) [ S_f\sin( \Delta m_{d}  \deltat' ) \nonumber\\  
- C_f\cos( \Delta m_{d}  \deltat' ) ] \Bigg\}  
\otimes R(\deltat-\deltat';\sigma_{\deltat}) , \qquad
\label{eq:pdf}
\end{eqnarray}
where the difference between the observed and true decay time differences $\deltat-\deltat'$ is described by
the empirical resolution function $R(\deltat-\deltat';\sigma_{\deltat})$.  This function is
parametrized as the sum of three Gaussians, a `core' and a `tail' Gaussian, each
with a width and mean proportional to $\sigma_{\deltat}$, and an outlier Gaussian centered at zero with a width of 8~ps.
The values of the \Bz\ lifetime $\tau_{\Bz}$ and the \Bz-\Bzb\ oscillation 
frequency $\deltamd$ are fixed to $(1.536 \pm 0.014)\ps$ and $(0.502 \pm 0.007)\ps^{-1}$ 
respectively~\cite{PDG2004}. 
We determine $S_f$ and $C_f$ separately for \DpDm, \DstpDm, and \DstmDp. 
If only tree-graph contributions are present, we expect
$S_{\DpDm}=-\stwob; C_{\DpDm}=0$, and $C_{\DstpDm}=-C_{\DstmDp}$. 
Additionally, under these conditions we have $S_{\DstpDm}=-X \sin(2\beta+\delta)$ 
and $S_{\DstmDp}=-X \sin(2\beta-\delta)$, with $X=\sqrt{1-C_{\DstmDp}^2}$ and where
$\delta$ is the difference of the strong phases for \BDstpDm\
and \BDstmDp. If the magnitudes of the amplitudes for \BDstpDm\ and \BDstmDp\ are equal~\cite{dstdexpect}, 
then $C_{\DstpDm}=C_{\DstmDp}=0$.
To determine the values of $\mistag$ for each of the tag categories and to increase the precision on the resolution function parameters,
we simultaneously fit to a large sample \bflav\ of reconstructed \Bz\ decays to the flavor 
eigenstates $D^{(*)-}h^+ (h^+=\pi^+,\rho^+$, and $a_1^+)$ and $\jpsi\Kstarz (\Kstarz\to\Kp\pim)$~\cite{babar-stwob-prd}.

The beam-energy substituted mass \mes $\equiv [(s/2 + \vec{p}_i \cdot \vec{p}_B)^2/E_i^2 - \vec{p}_B^{\;2}]^{1/2}$, where the initial total \epem 
four-momentum $(E_i,\vec{p}_i)$
and the $B$ momentum $\vec{p}_B$ are defined in the laboratory frame,
is used to determine the composition of the
reconstructed \DbothstpmDmp\ samples. We use only the region $\mes > 5.2$~\gevcc, which includes a large sideband of pure background 
events. These events are included in order to determine the properties of combinatoric background 
present in the signal region. 
Backgrounds are
incorporated with empirical descriptions of their \deltat spectra. The backgrounds include
prompt decays (associated with background from continuum events), and non-prompt decays with
a \deltat description similar to Eq.~\ref{eq:pdf}. Both components are convolved with a resolution function
distinct from that of the signal, parametrized as the sum of two Gaussians. 
Based on Monte Carlo studies we expect a significant flavor asymmetry in the non-prompt background of the
\BDstpmDmp\ samples, because the \Dstpm\ candidate is usually a true \Dstpm\, while the \Dpm\ is more often incorrectly reconstructed.
This flavor asymmetry is parametrized via values of $C_f$ and $S_f$ of the non-prompt background that are allowed to vary in the fit.

The \deltat and \mes\ distributions are fit simultaneously.
The \mes\ distribution, shown in Fig.~\ref{fig:mesplots}, allows a determination of a signal probability for each event.
In signal events, the values of \mes accumulate near the nominal $\BZ$ mass with a 
resolution of approximately 2.6~\mevcc. 
The fitted \mes shapes consist of a Gaussian distribution for the 
signal and an ARGUS function~\cite{Argus} for the combinatoric background.
The total number of selected candidates $N_{\rm cand}$ 
and the signal yield $N_{\rm sig}$ are shown in Table~\ref{tab:sample}. 
From detailed Monte Carlo simulations of 
generic $B$ decays, we expect some background events to peak in the $\mes$ signal region
due to cross-feed from other decay modes. The fraction of events in the signal Gaussian due to this peaking background is estimated
to be $(7.0 \pm 6.2)$\%  for \BDstpmDmp\ and  $(13.6 \pm 6.2)$\% for \BDpDm. 

\begin{table}[htb]
\begin{center}
\begin{tabular}{lrrc}
Sample 				& $N_{\rm cand}$  & $N_{\rm sig}$ & purity \\ \hline \hline
\BorBbarDstmDp				& 993	& $126 \pm 16$	& $0.49 \pm 0.03$ \\
\BorBbarDstpDm				& 1038	& $145 \pm 16$	& $0.49 \pm 0.03$ \\
\BorBbarDpDm				& 538	& $\;54\pm 11$	& $0.37 \pm 0.06$ \\
\end{tabular}
\end{center}
\caption{Candidates, signal yield and purity for each of the samples. 
The purity is defined as the fraction of signal events $N_{\rm sig}/N_{\rm cand}$ in the region $\mes > 5.27$~\gevcc.
}
\label{tab:sample}
\end{table}

The increase in statistics since our last measurement~\cite{oldbabarDstD} for \BDstpmDmp\ has allowed some refinements in the analysis.
These include an improved treatment of signal probabilities as determined from the \mes spectrum,
and additional floating parameters for the description of the background of the \CP sample.
We have also improved the event reconstruction, candidate selection, and tag-flavor determination.  The present effective tagging 
efficiency 
$Q = 30.5\%$~\cite{babar-stwob-newprl}, a relative increase of 5\% over the algorithm previously used. 

We perform separate fits for each of the three \CP\ samples. There are in total 54 floating parameters describing the 
\deltat distributions.
These are $C_f$ and $S_f$ for signal~(2) and background~(2), the average mistag fractions $\mistag_i$ and the
differences $\Delta\mistag_i$ between \Bz\ and \Bzb\ mistag fractions for each tag category $i$~(12), 
parameters for the signal \deltat resolution~(7), 
parameters for background \deltat\ distribution~(4) and resolution~(3) of the \bflav\ and \CP\ samples,
and values for  $\mistag_i$ and $\Delta\mistag_i$ for the prompt~(12) and non-prompt~(12) background of the \bflav\ sample.

The likelihood fits yield the following results:
\begin{eqnarray}
S_{\DstpDm} &=& -0.54 \pm 0.35{\rm (stat.)} \pm 0.07{\rm (syst.)}, \nonumber\\
C_{\DstpDm} &=& 0.09 \pm 0.25{\rm (stat.)} \pm 0.06{\rm (syst.)},\nonumber\\
S_{\DstmDp} &=& -0.29 \pm 0.33{\rm (stat.)} \pm 0.07{\rm (syst.)},\nonumber\\
C_{\DstmDp} &=& 0.17 \pm 0.24{\rm (stat.)} \pm 0.04{\rm (syst.)},\nonumber\\
S_{\DpDm} &=& -0.29 \pm 0.63{\rm (stat.)} \pm 0.06{\rm (syst.)}, \nonumber\\
C_{\DpDm} &=& 0.11 \pm 0.35{\rm (stat.)} \pm 0.06{\rm (syst.)}.  \nonumber
\end{eqnarray}

\begin{figure*}[tb]
\begin{center}
\includegraphics[width=0.32\linewidth]{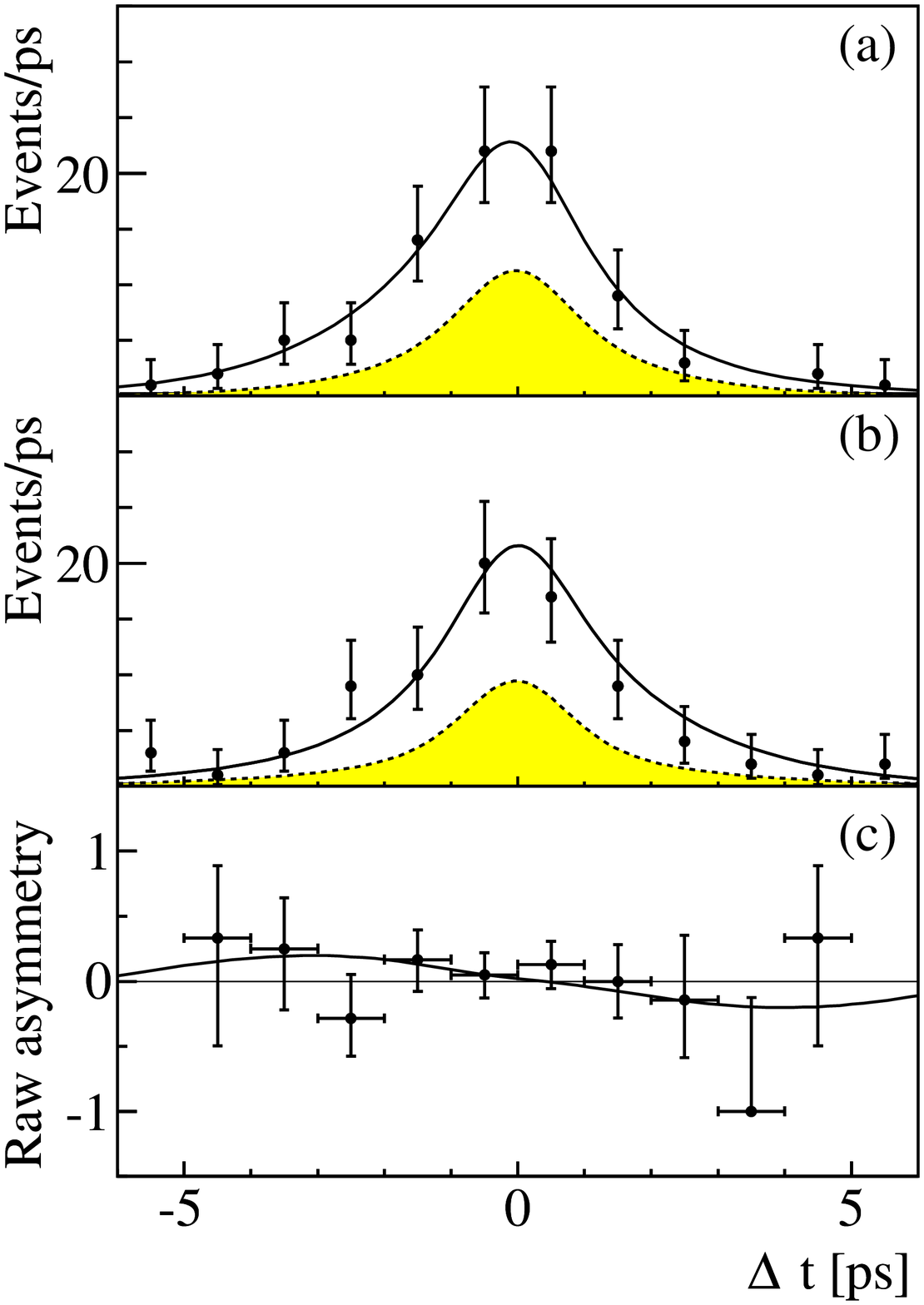}
\includegraphics[width=0.32\linewidth]{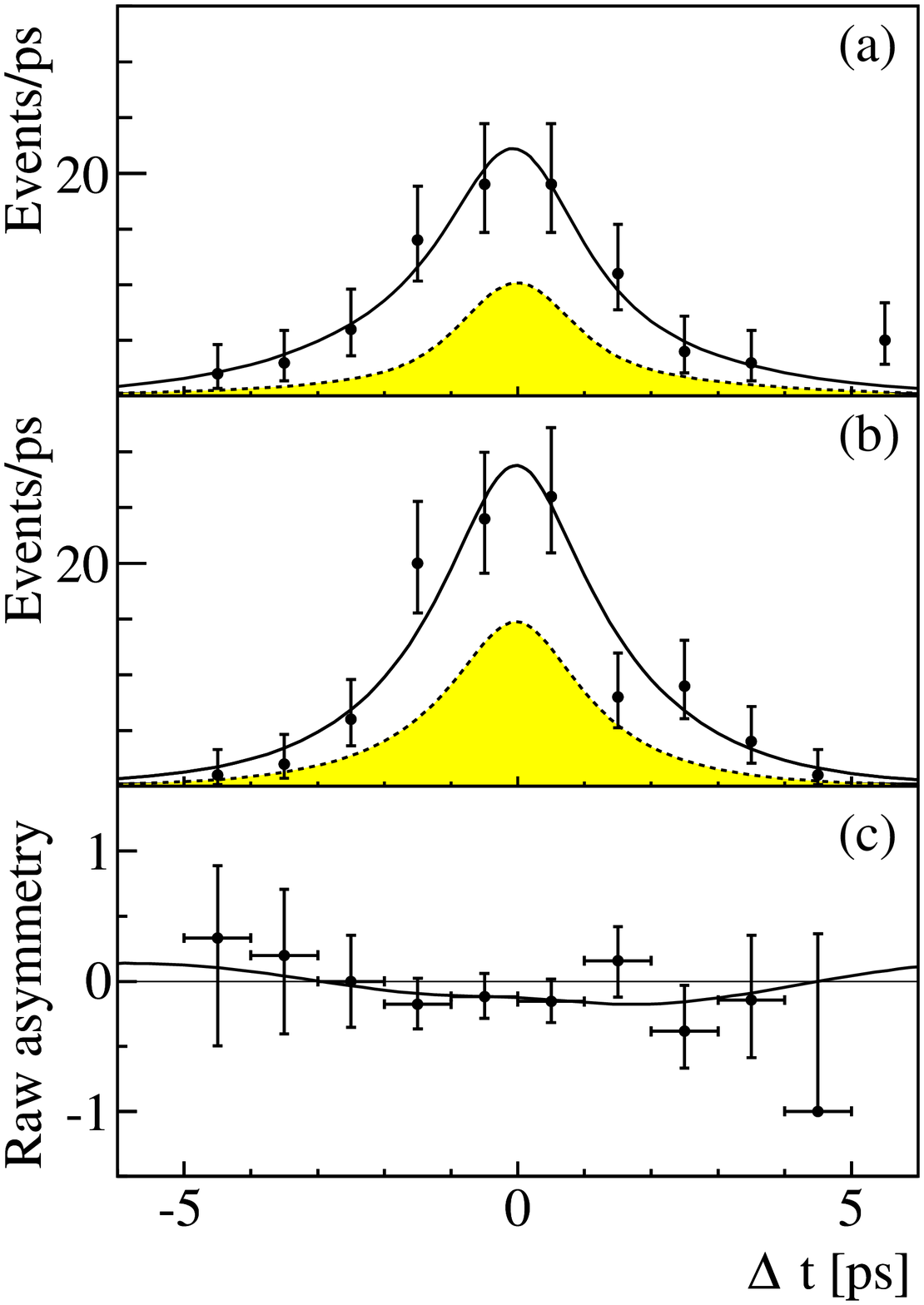}
\includegraphics[width=0.32\linewidth]{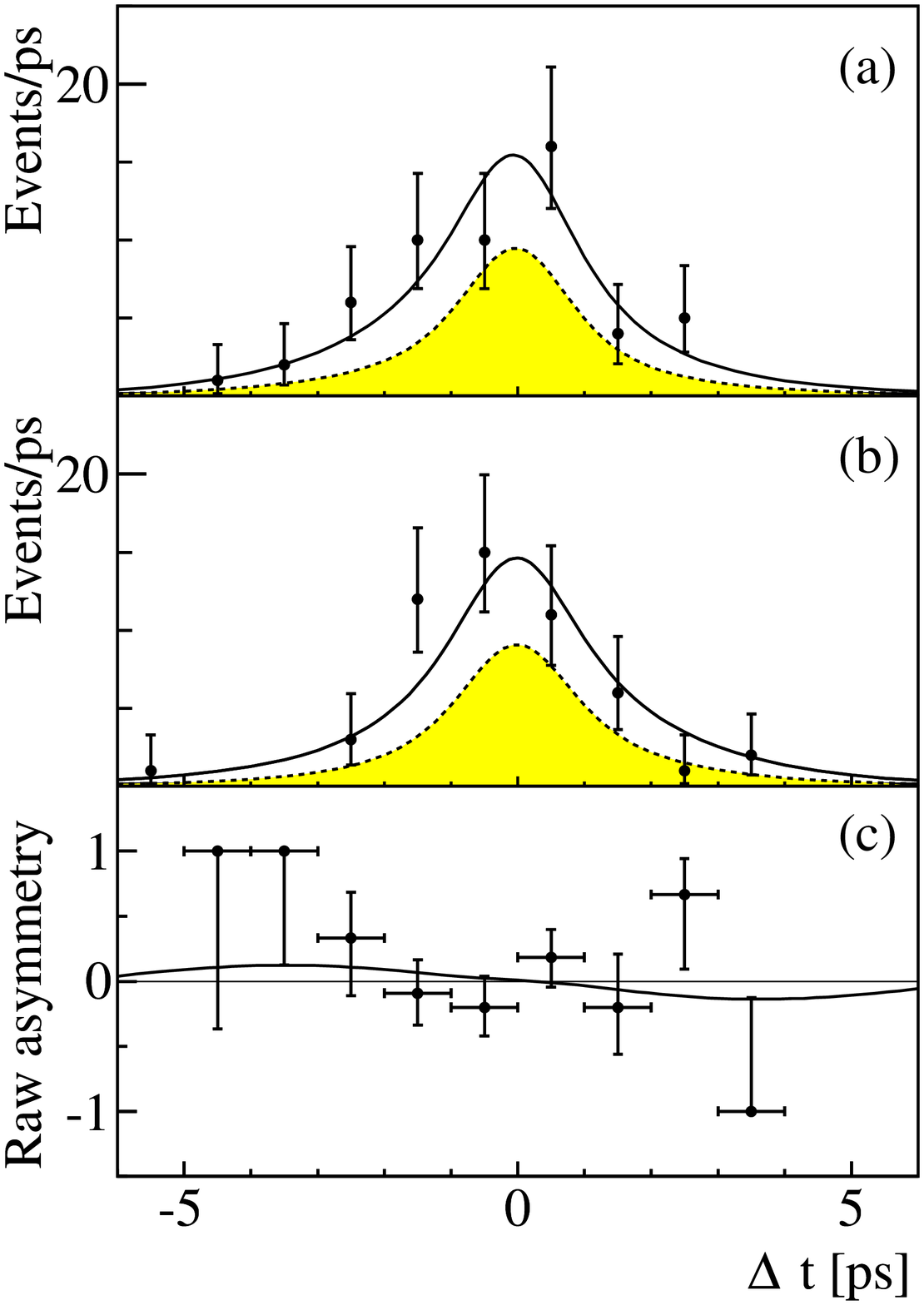}
\end{center}
\caption{Distribution of $\deltat$ and fit projections for \BorBbarDstpDm\ (left), \BorBbarDstmDp\ (middle) and  \BorBbarDpDm\ (right) 
candidates in the signal region $\mes > 5.27$~\gevcc with a \BZ\ tag (a) or a \BZBAR\ tag (b). The time-dependent \CP asymmetry is also shown (c).
The shaded areas represent the contributions from background events.
}
\label{fig:dtplots}
\end{figure*}

Projections of the fit onto \deltat\ for the three different \CP\ samples are 
shown in Fig.~\ref{fig:dtplots}, together with the raw \CP\ asymmetry 
\begin{equation}
A_{\CP}^{\rm raw} (\deltat) \equiv \frac{ {N}_+(\deltat)  -  {N}_-(\deltat) } { {N}_+(\deltat) + {N}_-(\deltat) },	
\end{equation}
where ${N}_+(\deltat)$ (${N}_-(\deltat)$) is the number of \BDbothstpmDmp\ events 
with a \BZ\ (\BZBAR) tag. 

The systematic uncertainties on $S_f$ and $C_f$ are separately evaluated for each of the decay modes.
The dominant systematic uncertainty is the precision to which we are able
to ascertain, using a Monte Carlo simulation, that the measurement method is unbiased (giving systematic uncertainties in the range 0.03-0.06).
Other important uncertainties are due to the amount of peaking background and its potential \CP\ asymmetry (0.01-0.02); assumptions 
on the \deltat\ resolution function (0.01-0.03); and potential differences between the mistag fractions for the \bflav\ and $B_{CP}$ 
samples (0.01-0.02).  Further sources of systematic uncertainty include the shape of the \mes\ distribution,
detector misalignment, uncertainty in the beam energies, and the possible interference between the 
suppressed $\bar b\to \bar u c \bar d$ amplitude with the favored $b\to c \bar u d$ amplitude 
for some tag-side decays~\cite{dcsd}. The total systematic uncertainty is considerably smaller than in our previous measurement 
(0.10-0.14), primarily due to fewer assumptions about the background of the \CP sample.

In summary, we have performed a first measurement of \CP asymmetries in the decay \BDpDm.
We have also updated our \CP asymmetry measurements in 
\BDstpDm\ and \BDstmDp, 
superseding our previously published results~\cite{oldbabarDstD}. 
The measured values are consistent with $S_f=-\stwob$ and $C_f=0$, expected in the SM for a tree-level-dominated 
transition with equal rates for \BDstpDm\ and \BDstmDp.
Since the dominant uncertainties are statistical, we anticipate improved precision with data collected in the future.

We are grateful for the excellent luminosity and machine conditions
provided by our \pep2\ colleagues, 
and for the substantial dedicated effort from
the computing organizations that support \babar.
The collaborating institutions wish to thank 
SLAC for its support and kind hospitality. 
This work is supported by
DOE
and NSF (USA),
NSERC (Canada),
IHEP (China),
CEA and
CNRS-IN2P3
(France),
BMBF and DFG
(Germany),
INFN (Italy),
FOM (The Netherlands),
NFR (Norway),
MIST (Russia), and
PPARC (United Kingdom). 
Individuals have received support from CONACyT (Mexico), A.~P.~Sloan Foundation, 
Research Corporation,
and Alexander von Humboldt Foundation.


\begin{thebibliography}{99}

\bibitem{CKM}
\hyphenation{Ko-ba-ya-shi}
N.~Cabibbo, \prl {\bf 10}, 531 (1963); M.~Kobayashi and T.~Maskawa, \progtp {\bf 49}, 652 (1973).

\bibitem{BCP}
A.B.~Carter and A.I.~Sanda, \prd {\bf 23}, 1567 (1981);
I.I.~Bigi   and A.I.~Sanda, \npb {\bf 193}, 85 (1981).

\bibitem{chargeconj}
Charge conjugate reactions are included implicitly unless otherwise noted.

\bibitem{babar-stwob-prl}
\babar\ Collaboration, B.\ Aubert {\em et al.},
\prl {\bf 89}, 201802 (2002).

\bibitem{belle-stwob-prl}
Belle Collaboration, K.\ Abe {\em et al.},
\prd {\bf 66}, 071102 (2002).

\bibitem{CKMfitters}
J.~Charles {\it et al.} 
  hep-ph/0406184; \\
  M.~Bona {\it et al.}  
  hep-ph/0501199.

\bibitem{dstdexpect}
Z.Z.~Xing, \prd {\bf 61}, 14010 (2000).

\bibitem{GrossWorah}
Y.~Grossman and M.~Worah, \plb{\bf 395}, 241 (1997).

\bibitem{DDgam}
A.~Datta and D.~London, \plb{584}, 81 (2004); 
J.~Albert, A.~Datta, and D.~London, \plb{605}, 335 (2005).

\bibitem{oldbabarDstD}
\babar\ Collaboration, B.\ Aubert {\em et al.},
\prl {\bf 90}, 221801 (2003).

\bibitem{belleDstD}
Belle Collaboration, T.\ Aushev {\em et al.},
\prl {\bf 93}, 201802 (2004).

\bibitem{babar-detector-nim}
\babar\ Collaboration, B.\ Aubert {\em et al.}, 
\nima{\bf 479}, 1 (2002).

\bibitem{PDG2004}
Particle Data Group, S.~Eidelman {\em et al.}, \plb {\bf 592}, 1 (2004).

\bibitem{FoxWolfram}
G.C.~Fox and S.~Wolfram, \prl{\bf 41}, 1581 (1978).

\bibitem{Fisher}
CLEO Collaboration, D.~Asner {\it et al.}, \jprd {\bf 53}, 1039 (1996).

\bibitem{Geant4}
{\tt GEANT4} Collaboration, S.~Agostinelli {\em et al.}, \nim{A 506}, 250 (2003).

\bibitem{babar-stwob-prd}
\babar\ Collaboration, B.\ Aubert {\em et al.}, 
\prd {\bf 66}, 032003 (2002).

\bibitem{babar-stwob-newprl}
\babar\ Collaboration, B.\ Aubert {\em et al.},
\prl {\bf 94}, 161803 (2005).

\bibitem{Gronau}
M.~Gronau, \plb{\bf 233}, 479 (1989).

\bibitem{Argus}
ARGUS Collaboration, H.~Albrecht {\em et al.}, \zpc{\bf 48}, 543 (1990).


\bibitem{dcsd}
O.~Long, M.~Baak, R.~N.~Cahn, and D.~Kirkby, \prd {\bf 68}, 034010 (2003).





\end{thebibliography}
\end{document}